# Probing Weyl Physics with One-dimensional Sonic Crystals

Xiying Fan,[1] Chunyin Qiu,[1,*] Yuanyuan Shen,[1] Hailong He,[1] Meng Xiao,[1] Manzhu Ke,[1] and Zhengyou Liu[1,2*]

[1]Key Laboratory of Artificial Micro- and Nano-Structures of Ministry of Education and School of Physics and Technology, Wuhan University, Wuhan 430072, China

[2]Institute for Advanced Studies, Wuhan University, Wuhan 430072, China

*Abstract.*—Recently, intense efforts have been devoted to realizing classical analogues of various topological phases of matter. In this Letter, we explore the intriguing Weyl physics by a simple one-dimensional sonic crystal, in which two extra structural parameters are combined to construct a synthetic three-dimensional space. Based on our underwater ultrasonic experiments, we have not only observed the synthetic Weyl points directly, but also probed the novel reflection phase singularity that connects inherently with the topological robustness of Weyl points. As a smoking gun evidence of the topological states of matter, the presence of nontrivial interface modes has been demonstrated further. All experimental data agree well with our full-wave simulations. As the first realization of topological acoustics in synthetic space, our study exhibits great potential of probing high-dimensional topological phenomena by such easily-fabricated and -detected low-dimension acoustic systems.

*Corresponding authors.
cyqiu@whu.edu.cn and zyliu@whu.edu.cn

*Introduction.*—Weyl semimetals, which are solid materials characterized by linearly crossing conduction and valence bands at isolated momenta, have attracted much attention in condensed matter physics [1-6]. The band crossing points—Weyl points (WPs), acting as monopoles of Berry flux in three-dimensional (3D) reciprocal space, are topologically robust against small perturbation. Weyl semimetals exhibit numerous peculiar properties, such as chiral anomaly [1-3] and topological Fermi arc surface states [4-6]. Recently, Weyl physics has been extended to artificial crystals that work for classical waves such as light [7-11] and sound [12-16]. Both the WPs [7,9,11,14,16] and associated surface arc states [8-11,14-16] have been observed successfully in such classical systems. Furthermore, exotic surface phenomena have been demonstrated through engineering equifrequency contours of the nontrivial surface states, such as topological negative refraction [15] and collimating transport of surface arc states [16]. Note that all the artificial Weyl crystals are real 3D structures, which involve notable complexities in fabricating sample and detecting signals. In this Letter, we explore the intriguing Weyl physics by using simple one-dimensional (1D) sonic crystals (SCs), in which the 3D space is synthesized by one physical dimension plus two additional structural parameters.

Fueled by the capability of realizing fascinating physics proposed in high-dimensional systems, constructing synthetic dimensions (i.e., extra degrees of freedom other than the physical one) has been widely used to relax experimental challenges encountered in real space [17-30]. Many approaches for creating synthetic dimensions have been introduced to various physical systems, such as cold atoms in optical lattices [17-21], photonic systems [22-28], and superconducting quantum circuits [29,30]. The development of synthetic dimensions even offers direct routes for probing the elusive physics of the systems beyond 3D [18-21,28], especially in exploring new topological phases of matter, in which dimensionality plays a crucial role [31,32].

By using our macroscopic experiments, we have demonstrated directly the presence of synthetic WPs, and confirmed the inherent vortex structure of reflection phases. The latter serves as a novel physical manifestation of the topologically robust

WPs. The consequent nontrivial interface states, accompanying huge local-field enhancements, have also been detected for distinct boundary truncations. The synthetic system exhibits some marked differences to real 3D Weyl crystals, due to the absence of periodicity in synthetic dimensions. For example, charge-neutrality is no longer indispensable [26]. Practical applications can be anticipated by exploiting the exotic phase singularity and strongly localized interface modes, such as to design exceptional acoustic sensors and highly directional acoustic antennas.

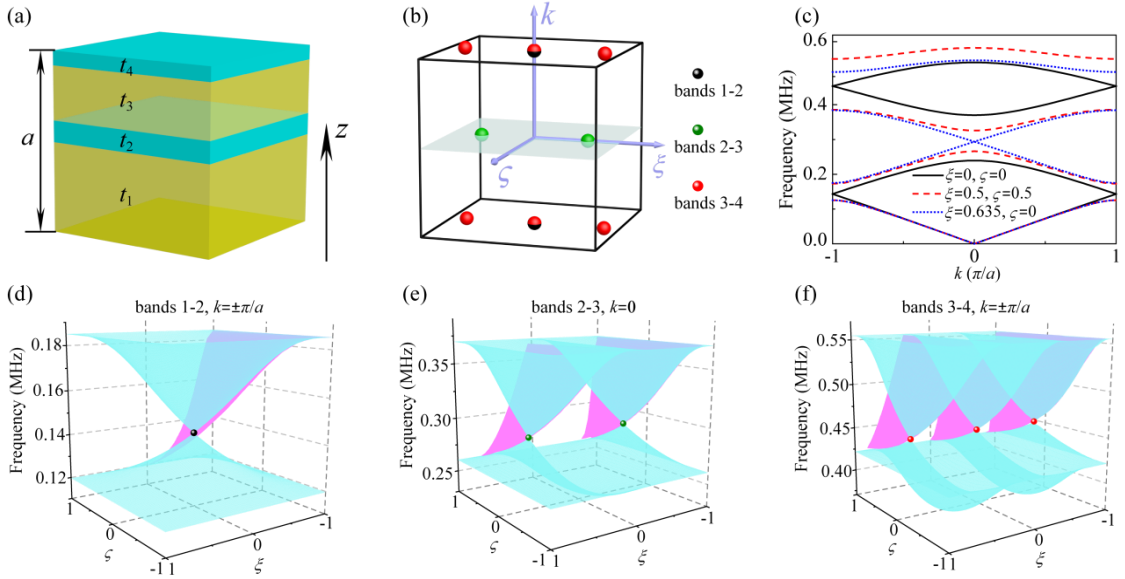

FIG. 1. WPs synthesized by the 1D layered SCs. (a) Four-layer-based unit cell, stacked alternately by water (yellow) and glass (blue) along the $z$ direction. (b) Virtual 3D Brillouin zone constructed by the 1D Bloch wavevector $k$ plus two structural parameters $\xi = (t_1 - t_3)/(t_1 + t_3)$ and $\varsigma = (t_2 - t_4)/(t_2 + t_4)$. The color spheres labeled the synthetic WPs formed between different neighboring bands [see (d)-(f)]. (c) Band structures for three SCs with different $\xi$ and $\varsigma$. (d)-(f) Bulk dispersions (cyan surfaces) plotted for the specified bands and wavevector in $\xi - \varsigma$ space. Pink surfaces indicate interface dispersions calculated with specific boundary conditions.

*Synthesized WPs by 1D SCs.*—As shown in Fig. 1(a), each unit cell of the 1D SC consists of two layers of water and two layers of glass [33], characterized by their thicknesses $t_1 \sim t_4$. The lattice constant $a = 5\,\mathrm{mm}$, and the total thicknesses of water and glass are $t_1 + t_3 = 4a/5$ and $t_2 + t_4 = a/5$, respectively. Two extra structural parameters, $\xi = (t_1 - t_3)/(t_1 + t_3)$ and $\varsigma = (t_2 - t_4)/(t_2 + t_4)$, are introduced to control all the layer thicknesses. Both parameters fall into the range of $[-1,1]$ and thus form a closed 2D space. Analog to 3D crystals, here a virtual 3D Brillouin zone [Fig. 1(b)] is constructed by the 1D Bloch wavevector $k$ plus the two parameters $\xi$ and $\varsigma$, in which Weyl physics can be explored.

We start from the special case $\xi = \varsigma = 0$, in which $t_1 = t_3$ and $t_2 = t_4$. The primitive lattice constant reduces to $a/2$ and the artificial band-folding creates linear degeneracy at $k = \pm\pi/a$ automatically [Fig. 1(c), black solid lines]. Here we focus on the lowest four bands. Band gaps are opened once leaving from $\xi = \varsigma = 0$, as exemplified by the SC with $\xi = \varsigma = 0.5$ (red dashed lines). Accidentally, two neighboring bands degenerate again at the high-symmetry momentum $k = 0$ or $k = \pm\pi/a$ if $\xi = 0$ or $\varsigma = 0$, as protected by mirror symmetry of the system. For example, an accidental linear degeneracy forms between the second and third bands at $k = 0$ for $\xi = 0.635$ and $\varsigma = 0$ (blue dotted lines). All above degenerate points are WPs in the synthesized 3D space. This is further verified by the conic intersections in the bulk dispersions plotted in $\xi - \varsigma$ space [Figs. 1(d)-1(f), cyan surfaces]. Besides those already indicated in Fig. 1(c), additional WPs are displayed clearly in the complementary band structures. Note that the symmetry of the system guarantees that [34], as long as there is one WP at $(\xi, \varsigma, k)$, there should be other ones at $(\pm\xi, \pm\varsigma, \pm k)$ with the same frequency, along with the same topological charge. Our further calculations state that all WPs [as exhibited in Fig. 1(b)] carry a unit charge of the same sign. Different from real Weyl crystals [35,36], here the total charge does not vanish since the synthetic parameter dimensions are not periodical. Interestingly, as predicted by the effective Hamiltonian model [26], the synthetic WPs always accompany reflection phase singularities in $\xi-\varsigma$ space, which in turn ensure the presence of topological interface states in truncated SCs [Figs. 1(d)-1(f), pink surfaces].

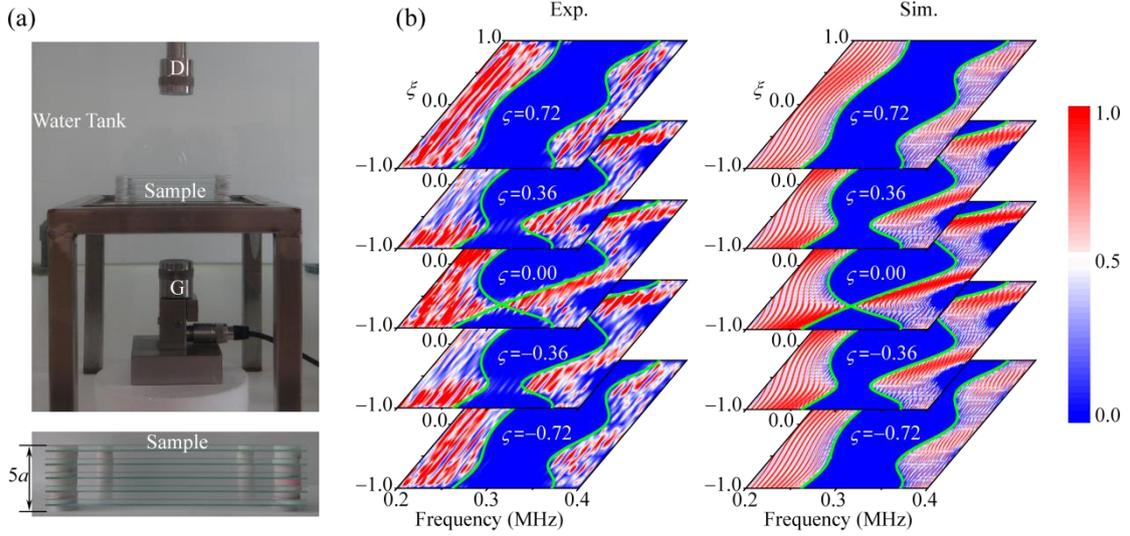

FIG. 2. Experimental evidence of synthetic WPs. (a) Experimental setup for transmission measurements. The bottom panel gives a zoom-in photo of the SC sample, where the thicknesses of water layers among the glass plates are controlled by solid spacers. (b) Measured and simulated transmission spectra through the samples with different structural parameters $\xi$ and $\varsigma$. Green lines indicate the numerical bulk dispersions for different $\varsigma$.

*Experimental observation of synthetic WPs.*—To experimentally detect the synthetic WPs, we have measured the ultrasonic transmission spectra through the 1D SCs with different structural parameters. Figure 2(a) shows our experiment setup. Each SC configuration consists of 5 spatial periods in total, stacked simply by 10 glass plates (of surface area $10\,\text{cm}\times 8\,\text{cm}$) with appropriate thicknesses and separations to attain desired $\xi$ and $\varsigma$. In each measurement, the SC sample is placed between two identical ultrasonic transducers, one for generating (G) and the other for detecting (D) sound signals, both with central frequency of 0.5 MHz and diameter of 2.5 cm. The transducers are placed far enough away from the sample to offer a good plane wave approximation, and meanwhile to separate the multi-scattering signals from them. The whole assembly is immersed in a big water tank. The transmission spectrum, defined by the ratio of the sound power transmitted through the sample to the incident signal measured without the sample in place, is attained through Fourier transform of the time-domain sound signal [37]. To map out the WPs we have considered totally 55 independent SC configurations (identified by 11 different $\xi$ and 5 different $\varsigma$). Here we focus on the frequency range around the two WPs touched by the second and third bands. Figure 2(b) shows the measured transmission spectra (left panel), compared with the corresponding numerical results (right panel). As expected, the frequency width of the measured transmission gap (indicated in blue) reduces and vanishes as $|\varsigma|$ approaches zero. The experimental data reproduce excellently the numerical ones, and exhibit clearly linear band crossings around the expected frequency and structural parameters. The experimental evidence for the three WPs formed by the third and fourth bands can be referred to *Supplemental Materials*, while the sole WP formed by the lowest two bands has not been captured precisely since its frequency goes beyond the optimal operation frequency of the transducers.

*Probing the reflection phase vortices and consequent interface states.*—As derived from the effective Weyl Hamiltonian, the nontrivial topology of the synthetic WPs enables inherently the presence of vortex-like reflection phase structures in the parameter space, in which the quantized phase winding numbers are identical to the corresponding topological charges of WPs [26]. Figure 3(a) presents a numerical example at 0.294 MHz, the frequency of WPs formed by the second and third bands. The phase distribution in $\xi-\varsigma$ space shows clearly two anticlockwise vortices (despite distorted) pinned at the projections of WPs (green spheres). The nontrivial phase distributions have been confirmed by our ultrasonic reflection experiments. Slightly different from the above transmission measurements, the reflected sound signal is recorded by the same transducer under the SC, from which the reflection phase $\varphi_{SC}$ can be extracted for each SC configuration. Figure 3(b) shows the phase distributions measured along the three anticlockwise parametric loops labeled in Fig. 3(a). Specifically, both the loops 1 and 2, involving 14 different SC configurations, encircle one WP, whereas the loop 3, involving 22 SC configurations, encloses no WP. As predicted from the simulations (lines), the measured phase profiles (circles) along the loops 1 and 2 exhibit an accumulative change of $2\pi$, in contrast to the case of loop 3 without apparent phase accumulation over it. Similar results for the higher Weyl frequency are provided in *Supplemental Materials*, which exhibit nontrivial phase accumulations along all the above parametric loops, consistent with the three WPs displayed in Fig. 1(f). The robust reflection phase singularity, as a novel manifestation of topologically stable WPs, has not been observed in real 3D Weyl crystals so far.

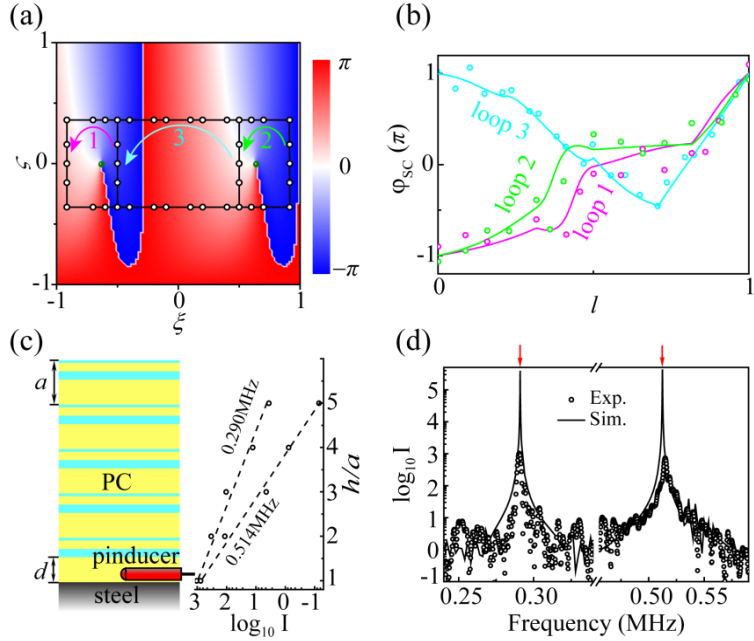

FIG. 3. Reflection phase singularity and the consequent topological interface states. (a) Reflection phases of the SCs with different structural parameters, simulated at the Weyl frequency 0.294 MHz. (b) Experimentally measured reflection phase distributions (circles) along the three closed parametric loops labeled in (a), capturing well the simulation results (lines). Here $l$ characterizes the discrete parameters along

the corresponding loop. (c) Left panel: Schematic illustration for the experimental detection of the interface states. Different from Fig. 2(a), now a steel substrate is used for emulating the hard boundary condition. Right panel: the energy density distributions inside the SC, detected at the two peak frequencies in (d). (d) Energy density spectra measured and simulated above the substrate, normalized by those measured without SC. The red arrows indicate the corresponding frequencies of interface states extracted from Figs. 1(e) and 1(f).

The vortex-like phase structure guarantees the existence of interface states if the SC is attached with a reflecting boundary. This conclusion follows the extended impedance argument [38,39]: the presence of interface states can be predicted by $\varphi_{SC} + \varphi_B = 2n\pi$, with $\varphi_B$ being the reflection phase from the hard boundary and $n$ being an integer. Since the reflection phase along any closed loop encircling one WP covers the full range of $2\pi$, the above phase condition can always be satisfied at appropriate structural parameters, regardless of the value of $\varphi_B$. Therefore, each WP is linked with one interface state deterministically, which serves as the bulk-boundary-correspondence interpretation of our synthetic Weyl system [26]. This has been numerically confirmed in Figs. 1(d)-1(f) by the interface dispersions (pink surfaces), which are simulated by imposing a rigid boundary condition at a distance $d = t_1$ to the lowest glass layer. In our experiments, as illustrated in Fig. 3(c), a steel plate of thickness 3.5 mm is placed under the SC to mimic the rigid boundary condition. (The transmission through the steel plate is as low as ~0.5% within the frequency window under consideration.) The ultrasonic signal is launched from the bottom and a pinducer (of diameter 1.5 mm) is inserted inside the water layer to act as sound receiver to detect the localized interface state. We consider a five-period SC slab with $\xi = 0.5$ and $\varsigma = 0.36$. Figure 3(d) shows the energy density spectrum (circles) measured at the lowest water layer and normalized by that measured without SC, comparing with the numerical one (solid line). Two notable peaks emerge at 0.290 MHz and 0.514 MHz, very close to those interface state frequencies (red arrows) extracted from Figs. 1(e) and 1(f). The exponential decay feature of such interface states has been further confirmed by detecting the sound intensity at different water layers [Fig. 3(c), right panel].

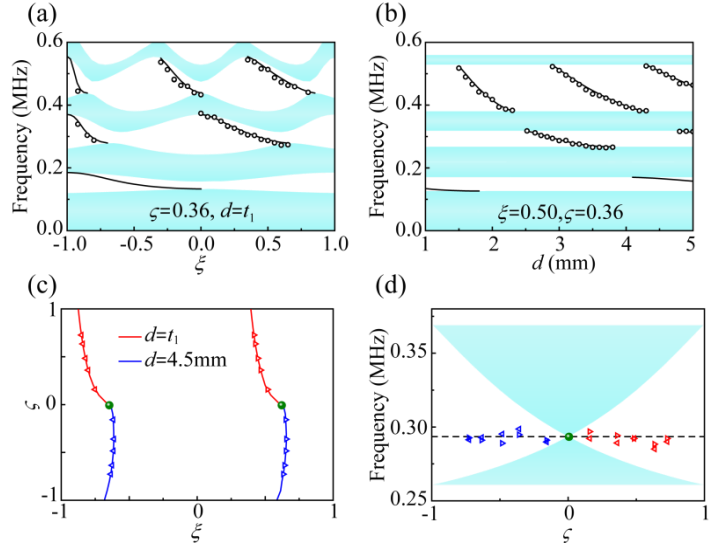

FIG. 4. Synthetic interface dispersions and acoustic Fermi arcs. (a) Frequencies of the interface states measured (circles) and simulated (lines) for different $\xi$ at fixed $\varsigma$ and $d$. The cyan regions are projected bulk bands. (b) Similar to (a), but for different $d$ at fixed $\xi$ and $\varsigma$. (c) Synthetic Fermi arcs at the Weyl frequency 0.294 MHz, simulated with two different boundary truncations. The green spheres label the projections of WPs. (d) Measured interface state frequencies for all the sample configurations labeled with the same symbols in (c).

Below we pay more attention to the topologically nontrivial interface states. Different geometric parameters are tailored and the frequencies of the interface states are extracted as above. Figure 4(a) presents the measured interface state frequencies for different $\xi$, which provides synthetic interface dispersions at the constant $\varsigma = 0.36$ for the specified boundary truncation $d = t_1$. The experimental data (circles) capture well the numerical results (lines). (The lowest interface band has not been detected since its frequency goes beyond the optimal operation frequency of the transducers.) Figure 4(b) provides the *d*-dependence of the interface state frequencies measured for a given SC, which shows that the frequencies can be tuned continuously by varying the distance between the SC and the substrate. As one of the most peculiar properties of 3D Weyl semimetals, the topological interface states possess open Fermi arcs that link the projections of oppositely charged WPs [4-6]. For comparison, here we explore the equifrequency contours of the interface states in $\xi-\varsigma$ space, i.e., synthetic acoustic Fermi arcs. Figure 4(c) presents the numerical results at the Weyl frequency 0.294 MHz, simulated for two different boundary truncations, $d = t_1$ and $d = 4.5\,\text{mm}$. It shows that each synthetic Fermi arc starts from one WP while terminates at the parametric boundary, and its shape is very sensitive to the SC boundary truncation. As an exotic physical manifestation, the two synthetic Fermi arcs (for any given boundary condition) exhibit no time-reversal connection, since $\xi$ and $\varsigma$ are not real Bloch momenta like periodical 3D Weyl crystals. To experimentally confirm the synthetic Fermi arcs, for each boundary truncation we consider 10 different SC configurations labeled on the arcs and extracted their interface state frequencies. As shown in Fig. 4(d), the measured results (red and blue triangles) are very close to the preset Weyl frequency (black dashed line).

*Potential applications.*—Many exciting applications could be stimulated based on this study. For example, the full $2\pi$ coverage of reflection phase can be used for sound wavefront shaping, considering a sample built by continuously varied structural parameters. The giant enhancement of sound energy density achieved by the strongly confined interface states [Fig. 3(d)] can be used to design acoustic sensors of high quality factor or devices that harvest sound energy. Particularly, in *Supplemental Materials* we show that such intriguing interface modes can also be combined to construct highly directional acoustic antennas [40-42]. Our simulation demonstrates that, the sound energy flow of a point source inserted inside the interface can be squeezed into a solid angle as narrow as 3.6$^\text{o}$. Together with the advantage of flexibly tunable operation frequency (by controlling $d$), such directional acoustic antennas are greatly useful for underwater communications and detections.

*Conclusions.*—We have experimentally explored the intriguing Weyl physics buried in the simple 1D SC platforms. A higher dimension can be constructed further by introducing other parameters (e.g., the total spatial filling ratio of glass), which enables a possible study of topological phenomena occurring at the dimensions inaccessible by real space [31,32]. This work may pave the way to explore the elusive high-dimensional topological physics by well-established low-dimensional SCs [27].

**Acknowledgements.**

This work is supported by the National Basic Research Program of China (Grant No. 2015CB755500); National Natural Science Foundation of China (Grant Nos. 11774275, 11674250, 11534013, 11747310); Natural Science Foundation of Hubei Province (Grant No. 2017CFA042).